\documentclass[preprint,11pt]{elsarticle}

\usepackage{lineno,hyperref}
\usepackage{amsmath}
\usepackage{amssymb}
\usepackage{amsbsy}
\usepackage{amsfonts}
\usepackage{amsopn}
\usepackage{amstext}
\usepackage{graphicx}
\usepackage{amssymb}
\usepackage{amsfonts}
\usepackage{amsmath}
\usepackage{graphicx}
\usepackage[english]{babel}
\usepackage{color}
\usepackage{slashed}
\usepackage{esint}
\usepackage[dvips]{epsfig}
\usepackage[dvips]{graphicx}
\usepackage{float}
\usepackage{units}
\usepackage{textcomp}

\usepackage{hyperref}
\usepackage{slashed}

\modulolinenumbers[5]

\journal{Physics Letters B}

\begin{document}

\begin{frontmatter}

\title{The bumblebee field excitations in a cosmological braneworld}

\author[pici]{L. A. Lessa}

\author[ufca]{J. E. G. Silva}

\author[pici]{C. A. S. Almeida}

\address[pici]{Universidade Federal do Cear\'a (UFC), Departamento de F\'isica, Campus do Pici, Fortaleza - CE, C.P. 6030, 60455-760 - Brazil}

\address[ufca]{Universidade Federal do Cariri(UFCA), Av. Tenente Raimundo Rocha, \\ Cidade Universit\'{a}ria, Juazeiro do Norte, Cear\'{a}, CEP 63048-080, Brazil}



\begin{abstract}
We investigated the effects of the spacetime curvature and extra dimensions on the excitations of a self-interacting vector field known as the bumblebee field. The self-interacting quadratic potential breaks the gauge invariance and the vacuum expectation value (VEV) of the bumblebee field $b_M$ violates the local particle Lorentz symmetry. By assuming the bumblebee field living in a $AdS_{5}$ bulk, we found an exponential suppression of the self-interacting constant $\lambda$ and the bumblebee VEV along the extra dimension. The fluctuations of the bumblebee upon the VEV can be decomposed into transverse and longitudinal modes with respect to $b_{M}$.  Despite the curvature, the transverse mode acquires massive Kaluza-Klein towers, while the longitudinal mode acquires LV mass $\lambda b^{2}$. On the other hand, the current conservation law prevents massive Kaluza-Klein modes for the longitudinal mode. For a spacelike $b_{M}$ along the extra dimension and assuming a FRW 3-brane embedded in the $AdS_{5}$ yields to an additional dissipative term to the longitudinal mode. The cosmological expansion leads to decay of the longitudinal mode in a time $\Delta t \approx H^{-1}$, where $H=\dot{a}/a$ is the Hubble parameter and $a(t)$ is the scale factor. For a timelike $b_{M}$, the longitudinal mode does not propagate on the brane and its amplitude decays in time with $a^{-3}$ and in the extra dimension with $z^{-\lambda b^{2}l^{2}}$.
\end{abstract}

\begin{keyword}
Spontaneous Lorentz symmetry breaking. Braneworld. Cosmology
\end{keyword}

\end{frontmatter}


\section{Introduction}

In recent decades, the possible Lorentz violating (LV) effects steaming from Planck scale has been extensively studied.
Some models in string theory \citep{KS}, very special relativity \citep{vsr}, noncommutative spacetime \citep{noncommutative} and loop quantum gravity \citep{lqg}, among others, enable Lorentz symmetry violation in the gravitational UV regime. A framework to explore Lorentz violating theories is provided by the Standard Model Extension (SME), wherein LV coefficients lead to violation of the particle Lorentz symmetry \cite{sme}.
A mechanism for the local Lorentz violating is provided by a spontaneous symmetry breaking potential due to self-interacting tensor fields. The vacuum expectation value (VEV) of these tensor fields yields to background tensor fields, which by coupling to the Standard Model (SM) fields violate the particle local Lorentz symmetry \cite{kostelecky,altschul,lessa}. Moreover, the spontaneous Lorentz violation allows the LV terms in the Lagrangian to satisfy the Bianchi identities, a key property for the gravitational field \cite{kostelecky}.

A self-interacting vector field, the so-called bumblebee $B_M$  has a VEV $b_M$ which defines a privileged direction in spacetime \cite{ks}. In flat spacetimes, causality and stability features of this model were studied, both classically \cite{ks2,bluhm,escobar} and at the quantum level \cite{hernaski,maluf}. The spontaneous breaking of the Lorentz symmetry leads to the emergence of Nambu-Goldstone (NG) modes and massive modes \citep{ks2}. For a quadratic potential, in the so-called Kostelecky-Samuel (KS) model in $3+1$ dimensions, the fluctuations around the vev $b_{M}$ yield to two transverse NG modes and one longitudinal massive mode. Since only the transverse modes are propagating, the photon can be interpreted as a NG mode of the bumblebee field instead of an elementary particle \cite{ks2,Seifert,seifert}.

In $3+1$ curved spacetimes, the modifications of the bumblebee upon the gravitational field were studied for black holes \cite{bertolami1,casana,malufbh}, wormholes \cite{ovgun} and cosmology \cite{capelo}. In higher dimensions, the bumblebee VEV modifies the Kaluza-Klein spectrum for bulk fields \cite{bertolami2,carroll,rizzo}. For a generalized bumblebee dynamics, an analysis of the fluctuations was performed in Ref.\cite{seifert}.

In this work, we are interested in study the propagation of the bumblebee fluctuations in curved spacetime. We consider the bumblebbe living in a five dimensional Anti de Sitter spacetime, $AdS_5$, with one spacelike extra dimension. Since $AdS_5$ is a maximally symmetric and conformal to a flat Minkowski spacetime, $AdS_5$ allows us to extended some results to curved spacetimes. We show that the bulk curvature makes the bumblebee self-coupling constant $\lambda$ depends on the spacelike extra dimension. Assuming two parallel 3-branes, this leads to an exponential suppression of $\lambda$, as in the Randall-Sundrum model \cite{rs1,rs2}. Assuming a homogeneous and isotropic Friedmann-Robertson-Walker (FRW) 3-brane embedded in $AdS_5$, for a spacelike VEV in the extra dimension the cosmological expansion produces a dissipative term for the longitudinal mode which decays in a rate $\Delta t \approx H^{-1}$. For a timelike VEV, the longitudinal fluctuation has an amplitude that vanishes as $a^{-3}$. These results reveal that additional modes steaming from spontaneous symmetry breaking of the Lorentz symmetry in the early universe may be suppressed by the cosmological expansion. That seems an expected feature since spontaneous violation of Lorentz symmetry is believed to occur during early universe phase transitions \cite{Bertolami1,kanno,Cheng}.

The work is organized as the following. In section \ref{sec2} we present the bumblebee dynamics in five dimensions, obtain the equations of motion for the fluctuations and study the propagation of these modes. In section \ref{sec3}, we investigate the effects of cosmic expansion considering a warped metric for both massless NG and massive modes. Final remarks are summarized in section \ref{sec4}. Throughout the text, we adopt the capital Roman indices ($A,B,... = 0,1,2,3,4$) denote 5-dimensional bulk spacetime indices, the Greek indices ($\mu , \nu , ... = 0,1,2,3$) the spacetime indices of the worldbrane.
Moreover, we adopt the metric signature $(-,+,+,+)$.

\section{Bumblebee dynamics in 5D} 
\label{sec2}

In this section, we consider the bumblebee field living in a 5D curved spacetime, called bulk, and see how the bulk curvature affects the dynamics of the bumblebee on a $3+1$ hypersurface called $3-$ brane. We start defining a 5D KS model action by \cite{ks,ks2}
\begin{equation} \label{action}
S = \int d^{5}x e \bigg[- \frac{\alpha}{4} B^{MN}B_{MN} -  \frac{\lambda}{2} (B^{M}B_{M} \pm b^2)^{2} \bigg],
\end{equation}
where, $b^2 = g^{MN}b_M b_N$ and the $e=\sqrt{-g}$ the determinant of the bulk metric in the five dimensional spacetime whose interval is $ds^{2}_5 = g_{MN}dx^{M}dx^{N}$. We consider a fixed background spacetime, i.e., the spacetime is not modified by the bumblebee.
Moreover, the  field-strength tensor $B_{MN}$ of the bumblebee field $B_M$ is defined as
$B_{MN} = \partial_{M}B_{N} - \partial_{N}B_{M}$.
In order to keep the bumblebee field with mass dimension one, we introduce the constant $\alpha$ with also mass dimension one, which we will discuss the details later.

The quadratic potential chosen induces the spontaneous Lorentz violation, where $\lambda$ is a mass dimension one positive self-interaction coupling constant, $b^{2}$ is a positive constant with squared mass dimension and the $\pm$ sign meaning if $b_{M}$ is spacelike or timelike. Moreover, the vacuum condition $V=0$ implies the existence of a vacuum expectation value $<B_M>=b_M$ is the form
\begin{equation} \label{norma}
g^{MN}b_{M}b_{N} = \mp b^{2}.
\end{equation}


In order to investigate the effects of spacetime curvature and extra dimensions on the bumblebee fluctuations, we adopt a special warped geometry in the form \cite{rs1,rs2}
\begin{equation} \label{metric1}
ds^{2}_5 = e^{-2cy}ds^{2}_{brane}+dy^{2} ,
\end{equation}
where $e^{-2c y}$ is the so-called warp factor of the Randall-Sundrum model, which depends only on the fifth dimension $ y $. For a flat 3-brane, i.e., $ds^{2}_{brane}=\eta_{\mu\nu}dx^\mu dx^\nu$, this metric describes an Anti De Sitter spacetime, $AdS_5$, which in the conformal coordinate $z=\frac{e^{cy}}{c}$ takes the form \cite{rs1}
\begin{equation} \label{adsmetric}
ds^{2}_5 =  \frac{l^2}{z^2}(\eta_{\mu\nu}dx^\mu dx^\nu +dz^2) ,
\end{equation}
where $l=1/c$ is the $AdS$ radius and
\begin{equation}
    z=l e^{cy}
\end{equation}
is the conformal coordinate. It is worthwhile to mention that this five-dimensional line element preserves four-dimensional Poincaré invariance of the 3-brane embedded in the $AdS_5$ bulk. The $AdS_5$ is a maximally symmetric spacetime, i.e., $ R_{MNPQ}= \frac{R}{20}(g_{NQ}g_{MP} - g_{NP}g_{MQ})$, 
where $R=-20/l^2$ is the $AdS_5$ constant and negative Ricci scalar. The Anti de Sitter spacetime is a solution of Einstein equation with a negative cosmological constant of form
$R_{MN} -\frac{R}{2}g_{MN}+ \Lambda g_{MN}=0$, with $\Lambda=-6c^2$. It is upon this symmetric background spacetime that we study the behaviour of the bumblebee fluctuations.

Before approaching the equation of the motion (EoM) for the bumblebee field and the propagation of the fluctuating modes that appear in the KS theory in five dimensions, let us first analyze the effects of bulk curvature effective action in $3+1$ dimensions and the corresponding effective constants $ \lambda$ and $\alpha$.
Suppose that the bumblebee field and its VEV have a dependence on the conformal extra dimension $z$ of the form
$B_{M} = \tilde{B}_{M}(x^{\mu})\Upsilon(z)$ and $b_{M} = \tilde{b}_{M}(x^{\mu})\Psi(z)$.
Thus, the VEV condition (\ref{norma}) leads to
\begin{equation}
b_{M} = (l/z)\tilde{b}_{M}(x^{\mu})
\end{equation}
where $\tilde{b}_{M} \tilde{b}^{M} = \tilde{b}^2$ is constant with respect to the flat 5-D Minkowski metric $\eta_{MN}dx^M dx^N =\eta_{\mu\nu}dx^\mu dx^\nu +dz^2$. Supposing that the bumblebee field decays as the VEV $b_M$, we obtain $B_{M} =  (l/z) \tilde{B}_{M}(x^{\mu})$.

Let us now consider two parallel and fixed 3-branes, one at the origin and other at $y=L$, the well-known RS-I model\cite{rs1}.
For $b_{M} = (l/z)\tilde{b}_{M}(x^{\mu})$ and $B_{M} = (l/z)\tilde{B}_{M}(x^{\mu})$, the potential term leads to the y-dependent self-interacting coupling constant $\lambda_{eff}=(\frac{l}{z})^5$, or in the y coordinate,
\begin{equation}
    \lambda_{eff}=\lambda_{eff}(y)=\lambda e^{-5cy}.
\end{equation}
Therefore, the $AdS_5$ curvature in the RS I model yields to an exponential suppression of the bumblebee self-interaction constant between the 3-brane at $y=0$ and at $y=L$. Note that, although the bumblebee vev $b_M$ decays as $b_M = e^{-cy}\tilde{b}_M$, the $b^2$ is kept constant throughout the entire $AdS_5$.

In its turns, $\alpha$ varies with the extra dimension as
\begin{equation}
    \alpha_{eff}=\alpha(y)=\alpha e^{-2cy}.
\end{equation}
Thus, at the visible brane at $y=L$, the Lorentz violating effects given by the self-interacting are much more suppressed than those described by the usual kinetic term.

In the RS II model, wherein there is only one 3-brane at the origin \cite{rs2}, by integrating out the 5-D potential term in the extra dimension yields to
\begin{equation}
    S_V = -\lambda\int_{0}^{\infty}(l/z)^5 dz\int{d^4 x \sqrt{-g_4}(\eta^{\mu\nu}}\tilde{B}_\mu \tilde{B}_\mu \pm b^2)^2.
\end{equation}
Thus, the effective (3+1) coupling constant in the brane at $y=L$ is given by
\begin{equation}
\lambda_{eff} = \frac{1}{2c}\lambda.
\end{equation}
Since $c$ has mass dimension one, for a five dimensional bumblebee self coupling constant $\lambda$ with mass dimension one, then $\lambda_{eff}
$ is dimensionless. For the constant $\alpha$, integrating the kinetic term along the extra dimension, we obtain
\begin{equation}
     S_K = -2\alpha\int_{0}^{L}(e^{-2cy}dy)\int{d^4 x B^{\mu\nu}B_{\mu\nu}}.
\end{equation}
Accordingly, the relation between the five dimensional constant $\alpha$ and the four dimensional $\alpha_{eff}$ is given by
\begin{equation}
\alpha_{eff}= \frac{\alpha}{c}.   
\end{equation}
Once again, the effective constants depend not on the length of the extra dimension but on the $AdS_5$ spacetime curvature. A similar dimensional reduction result appears in the original RS II model for the gravitational constant \cite{rs2}.

\subsection{Equations of motion for the fluctuations}
In this part of the work, we will develop the equations of motion for bumblebee fluctuations considering that $ds^{2}_{brane}$ is curved. Varying with respect to the bumblebee field the action \eqref{action}, we obtain the equations of motion \cite{ks,ks2}
\begin{equation} \label{eqmov}
D_{N}B^{NM} = J^{M}_{B}
\end{equation} 
where $J^{M}_{B}$ arises from the bumblebee self-interaction and it is given by \cite{ks,ks2}
\begin{equation}
J^{M}_{B} = 2 V'B^{M}.
\end{equation}
Moreover, the antisymmetry of the bumblebee field strength $B_{MN}$ 
implies a conservation law:
\begin{equation} \label{vinculo}
D_{M} J^{M}_{B} = 0.
\end{equation}
Now consider the fluctuation about the bumblebee VEV, i.e.,
\begin{equation} \label{flutuação}
B_{M} \approx b_{M} + \chi _{M},
\end{equation}
where $<B_{M}> = b_{M}$. The linearized Lagrange density takes the form
\begin{equation} \label{lagrangiana1} 
e\mathcal{\tilde{L}}_{KS} =  - \frac{1}{4} e b^{MN}b_{MN} - \frac{1}{4} e \chi^{MN} \chi _{MN} - \frac{1}{2} e\chi^{MN} b_{MN} -2e\lambda (b_{M} \chi ^{M}) ^{2}, 
\end{equation} 
where $b_{MN} = \partial_{M} b_{N} -  \partial_{N} b_{M}$ and $\chi_{MN} = \partial_{M}  \chi_{N} -  \partial_{N}  \chi_{M}$.
Thus, the equation of motion for the fluctuations is given by
\begin{equation}
\label{fluctuationeom}
\Box \chi^{N} - D^{N}(D_{M}\chi^{M}) - R_{T}^{\ N}\chi^{T} + D_{M}b^{MN} \approx 4\lambda (\chi^{M}b_{M})b^{N},
\end{equation}
where $\Box=D_{M}D^{M}=g^{MN}D_M D_N$ is the 5D D'Alembertian operator and $R_{MN}=R^{P} \ _{MPN}$ is the Ricci tensor in 5D.

It is worth noting that for the vacuum solution, i.e., $B_{M} = b_{M}$ the Eq. (\ref{eqmov}) is given by $D_{M}b^{MN}=0$, since we have a minimum of the potential, $V'=0$, for the vacuum solution. But as shown in Eq.(\ref{fluctuationeom}), when we assume fluctuations around the vacuum value, this equation of motion is modified by the fluctuations.

The Eq.(\ref{fluctuationeom}) has a similar form of the fluctuations EoM in flat spacetime \cite{ks2}, except for the covariant derivatives, the coupling to the Ricci tensor and the varying VEV. It is worthwhile to mention some interesting features of the fluctuations EoM in general curved spacetimes, before we focus on the specific braneworld scenario. For instance, assuming that the background spacetime geometry is a vacuum, the Ricci tensor vanishes identically, i.e., $R_{MN}=0$. Thus, the third term in Eq.\eqref{fluctuationeom} vanishes not only in Minkowsky spacetime but also in any  background spacetime vacuum. For a vacuum  maximally symmetric spacetime, $R_{MN}=\frac{R}{3}g_{MN}$, and thus the third term in Eq.\eqref{fluctuationeom} provides a mass term for the fluctuation field $\chi_M$.

Unlike the flat spacetime, which allows us to define a constant background VEV, $\partial_M b_N =0$, the curvature constrains the $b_M$ VEV. In fact, assuming a covariant constant $b_M$, i.e., $D_M b_N=0$, leads to the constrain $b_{M}R^{M } \ _{NPQ}=0$. This constrain means that the curvature vanishes in the direction of the background vector.If we adopt a less restrictive VEV definition, by assuming that the VEV norm $b^2=g^{MN}b_M b_N$ is constant, the VEV satisfies
\begin{equation}
(D_{N}b^{M})b_{M} = 0.
\end{equation}

Since the VEV defines a preferred direction in spacetime, we can decompose $\chi_{M}$ into transverse $A_M$ and longitudinal $\beta$ modes with respect to $b_{M}$ \cite{ks2} \begin{equation} \label{decomp}
\chi _{M} = A_{M} + \beta \hat{b}_{M},
\end{equation}
where by defining the projection operators $P^{||}_{MN} = \frac{b_{M} b_{N}}{b^{A}b_{A}}$ and $P^{\perp}_{MN} = g_{MN} - \frac{b_{M} b_{N}}{b^{A}b_{A}}$,
we have $A_{M} = P^{\perp}_{MN} \chi^{N}$ and $\beta\hat{b}_{M} = P^{||}_{MN} \chi^{N}$.
As result, we have to $A_{M}b^{M}\approx0$ and $\hat{b}_{M}\hat{b}^{M} = \mp1$, where $\hat{b}_{M} = \frac{b_{M}}{\sqrt{b^{2}}}$.  Using the decomposition [\ref{decomp}], the smooth quadratic potential term becomes
\begin{equation}
V \approx 4 \lambda[(\hat{b} ^{A}b_{A})\beta]^{2}
\end{equation}
i.e., $V(X)\neq0$, therefore the $\beta$ is the longitudinal mode.  Before this linearized bumblebee current, we have Eq. (\ref{vinculo}) the linearized conservation law
\begin{equation} \label{law}
D_{M}(\beta b^{M}) \approx 0.
\end{equation}
Using the decomposition (\ref{decomp}) and the conditions $b^2$ constant, \eqref{law} and $A^{M}b_{M}=0$, the equation of motion for the longitudinal mode $\beta$ is given by 
\begin{align} \nonumber \label{motionmode}
&(\Box \beta)(\hat{b} ^{M}b_{M}) - [ R_{N} \ ^{M} \hat{b} ^{N}b_{M} - (\Box \hat{b}_{M})b^{M} + 4\lambda (\hat{b} ^{M}b_{M})(b ^{N}b_{N}  )] \beta \approx\\
& (D_{N} b^{MN})b_{M}  + b_{M}[D^{M}(D_{N}A^{N})] + R_{T} \ ^{M}A^{T}b_{M} + 2(D_{N}A^{M})(D^{N}b_{M})\nonumber \\
&+ A^{M}(D^{N}D_{N}b_{M}),
\end{align}
while the transversal mode $A_M$ is governed by
\begin{eqnarray} \label{massless}
    \Box A^{N} - D^{N}D_{M}A^{M} - R _{M} \ ^{N}A^{M} &\approx&  [4\lambda(b^{M}\hat{b}_{M})b^{N}-R_{M} \ ^{N}\hat{b}^{M}]\beta\nonumber\\
    &+& \Box (\beta\hat{b}^{N}) + D_{M}b^{NM} .
\end{eqnarray}
Unlike in the Minkowski spacetime \cite{ks2}, in a general spacetime it is not possible to decouple the longitudinal and transverse modes.  This is due to the curved spacetime nature, which leads to a varying VEV and new couplings between the fluctuations and the curvature tensor.

Since $AdS_5$ is a maximally symmetric spacetime, $R_{MN} = \frac{R}{D}g_{MN}$, and thus, the term $ R_{N} \ ^{M}A^{N}b_{M}$ in Eq.\eqref{motionmode} vanishes.
For $g_{MN}=(l/z)^{2}\eta _{MN}$ and $b_{M}=(l/z) \tilde{b}_{M}$,
where $\tilde{b}_{M}$ is a constant, we notice that the last two terms in (\ref{motionmode}) also vanish. Eq.\eqref{fluctuationeom} also simplifies, for the Ricci coupling term in Eq.\eqref{fluctuationeom} leads to the non-massive terms $R_{MN}b^M b^N= \frac{R}{(D)}b^2$ and $R_{MN}A^M b^N =0$. 

\section{The KS model on a cosmological background} \label{sec3}

Since the curvature strongly couples the longitudinal and transverse modes, let us consider the propagation of the bumblebee fluctuations on a rather symmetric spacetime. Thus, consider a 3-brane geometry described by the homogeneous and isotropic Friedmann-Robertson-Walker metric (FRW)
\begin{equation} \label{metric}
ds^2_{brane} = - dt^2 + a(t)^2 \bigg[ (dx^{1})^2 + (dx^{2})^2 + (dx^{3})^2   \bigg],
\end{equation} 
where $a(t)$ is the scale factor.


\subsection{Spacelike vev}

Consider a spacelike VEV in the conformal coordinates with only a nonvanishing fifth component in the form
\begin{equation} \label{vev4}
b_{M} = (0,\vec{0},\tilde{b}\left(l/z\right)),
\end{equation}
where $\tilde{b}$ is a constant that arises from the constant norm condition (\ref{norma}). The VEV choice in \eqref{vev4} has a vanishing field strength, i.e., $b_{MN}=0$. In addition, this VEV choice constrains the transverse mode $A^{M}$ to the 3-brane, i.e., $ A^{4}= 0 $.

The linearized Lagrangian for this spacelike VEV is given by
\begin{eqnarray} \label{ac}
e\mathcal{\tilde{L}}_{KS} &\approx&  - \frac{1}{4} e F^{MN} F _{MN} - \frac{1}{2}e (\partial_{M}\beta)(\partial^{M}\beta)(\hat{b}^{N}\hat{b}_{N})+ \frac{1}{2}e (\partial_{M}\beta)(\partial^{N}\beta)(\hat{b}^{M}\hat{b}_{N})\nonumber\\
&-&eF^{MN}(\partial_{M}\beta)\hat{b}_{N}-2e\lambda (\beta \hat{b}_{M} b^{M}) ^{2}  \nonumber\\
& \approx &   - \frac{1}{4} e F^{MN} F _{MN} - \frac{1}{2}e (\partial_{\mu}\beta)(\partial^{\mu}\beta) + \hat{b}^{4}(\partial_{4}A_{\mu})(\partial^{\mu}\beta) - 2e\lambda\tilde{b}^{2}\beta^{2}.
\end{eqnarray}As we can see, the only term responsible for the coupling between the longitudinal and transverse mode is $\hat{b}^{4}(\partial_{4}A_{\mu})(\partial^{\mu}\beta)$. 

In order to analyse the coupling between the modes, as well as their dependence on the extra dimensions, 
let us perform the Kaluza-Klein (KK) decomposition. For the transverse mode, let us search for solutions of the form  $A_{\mu}(x,z)=\tilde{A}_{\mu}(x)\Gamma(z)$. 
From Eq. (\ref{massless}), the brane dependence of the transverse mode $\tilde{A}_\mu$ satisfies
\begin{equation}\label{kk1}
    \frac{1}{\sqrt{-g_{4}}}\partial _{\mu}(\sqrt{-g_{4}}g^{\mu\alpha}_{4}g^{\nu\lambda}_{4}\tilde{F}_{\alpha\lambda}) = m^{2}\tilde{A}^{\nu},
\end{equation} 
whereas the extra dimension dependence is governed by
\begin{equation}\label{kk2}
    \Gamma '' - \frac{1}{z}\Gamma'+ m^{2}\Gamma = 0,
\end{equation}
where $\tilde{F}_{\alpha\lambda}= \partial_{\alpha}\tilde{A}_{\lambda}-\partial_{\lambda}\tilde{A}_{\alpha}$ and the $m$ is a constant called KK mass.  This constant can be arbitrarily small and it runs in the range $- \infty < m < \infty$ . The solution of Eq. (\ref{kk2}) are the Bessel functions of the first and second kind, respectively, given by $\Gamma(z) = \Gamma_{1}z J_{1}(mz) + \Gamma_{2}z Y_{1}(mz)$, where $\Gamma_{1,2}$ are constants. Thus, likewise the gauge vector field, the bumblebee transverse mode acquires a mass due to the dimensional reduction. For the massless mode, i.e., $m^{2}=0$,  the Eq. (\ref{kk2}) leads to a solution $\Gamma(z) = \Gamma_{0} + \frac{c_1}{2}z^2$, which grows with $z$., whereas $\tilde{A}_\mu$ satisfies $D_{\mu}\tilde{F}^{\mu\nu}=0$.

Note that the corresponding term $\hat{b}^{4}(\partial_{4}A_{\mu})(\partial^{\mu}\beta)$ of the action does not appear in EoM neither in (\ref{kk1}) nor in (\ref{kk2}). This follows from the $A^{M}b_{M}=0$ condition. We can see this better by choosing $N=\nu$ in Eq.(\ref{massless}). Note that all terms on the right side are zero. This coupling term will appear in the EoM of the longitudinal mode, as we will see below.

For the longitudinal mode, by assuming the KK decomposition $\beta (x,z)= \tilde{\beta}(x)\Upsilon(z)$, the Eq.\eqref{motionmode} simplifies  into
\begin{equation} \label{eqkk}
\bigg[D_{4}D^{4}\Upsilon +\bigg( \frac{4}{l^{2}} + \hat{b}^4\Box \hat{b}_{4}\bigg) \Upsilon \bigg]\tilde{\beta} +  \bigg[D_{\mu}D^{\mu}\tilde{\beta} + 4\lambda\tilde{b}^{2}\tilde{\beta}\bigg] \Upsilon = \hat{b}_{4}D^{4}D_{\mu}A^{\mu}.
\end{equation}
Considering the conservation law Eq.(\ref{law}) for the spacelike case, we find that
\begin{equation} \label{law1}
    \frac{\tilde{b}\tilde{\beta}}{(l/z)^{5}}\partial_{4}\bigg((l/z)^{4} \Upsilon \bigg) = 0,
\end{equation}
i.e,  the solution is given by $\Upsilon(z)=\Upsilon_{0}\left(z/l\right)^4$, where $\Upsilon_{0}$ is a constant. Substituting $\Upsilon(z)$ in the equation above, we find that
\begin{equation}
 \label{longitudinalmodeonbranespace}
         D_{\mu}D^{\mu}\tilde{\beta} + 4\lambda\tilde{b}^{2}\tilde{\beta} = \Upsilon_{0} (l/z)^{3} \partial_{4}\Gamma D_{\mu}\tilde{A}^{\mu}
         \end{equation}
Note that the first term of Eq.(\ref{eqkk}), which carries the dependency with the extra dimension, vanishes due to Eq.(\ref{law1}). Moreover the KK mass of the transverse mode couples $\tilde{A}_\mu$ to $\tilde{\beta}$. For massless transverse mode, $m^2=0$, and considering that field $\Gamma$ vanishes at infinity, i.e., $\Gamma=\Gamma_{0}=const.$ ,  the transverse and longitudinal modes decouple, hence the U(1) symmetry is recovered. Another important point that we need to emphasize from Eq.(\ref{longitudinalmodeonbranespace}) is that due to the current conservation law (\ref{law1}), the longitudinal mode in the spacelike case did not generate Kaluza-Klein towers. An analysis of KK towers in presence of Lorentz-violating aether fields in space-time  with extra dimensions was done in Ref.\cite{carroll}
 
 Likewise the gauge vector field that is not normalized in the effective action in (1+3)-dimensions (RS-II), the bumblebee transverse fluctuations also diverges in action. This similarity is due to the fact that the two models share the same kinetic term, $F_{MN}F^{MN}$. The kinetic term in action for the zero mode solution is $S_{k} \backsim \Gamma_{0}^{2} \int dz (l/z) \int d^{4}x \sqrt{-g} \tilde{F}_{\alpha\lambda}\tilde{F}^{\alpha\lambda}$, where $\Gamma_{0}$ is a constant. Indeed, if the warp factor is factorized out of
the effective action, one obtains non-normalizable solutions from the equations of motion to the gauge field.

In order to localize the gauge field, the authors of \cite{dilaton} introduced a scalar field called dilaton $\pi(z)$ which couples to the kinetic term of $A_{\mu}$ field
and leads to the localization.  We can achieve a brane localized massless transverse mode by considering that the parameter $\alpha$ introduced in Eq.(\ref{action}) depends on the extra dimension as $\alpha = e^{-\frac{\zeta \pi (z)}{2}}$, where $\zeta$ is a dimensionless dilaton coupling depending on the details of the underlying theory. Therefore the normalization of the transverse field is dictated by the dynamics of the dilaton field in the following way $S_{\perp}= - \frac{1}{4}  \Gamma_{0}^{2}\int  e^{-\frac{\zeta \pi (z)}{2}} (l/z) dz \int d^{4}x \sqrt{-g_{4}} \tilde{F}^{\mu\nu}\tilde{F}_{\mu\nu}$ . And finally, we have that the longitudinal field also has its normalization controlled by $\pi(z)$, since the longitudinal part of the action with $m=0$ is given by $S_{\parallel}= - \frac{1}{2} \int e^{-\frac{\zeta \pi (z)}{2}} (z/l)^{5}dz  \int d^{4}x \sqrt{-g_{4}}\partial_{\mu}\tilde{\beta}\partial^{\mu}\tilde{\beta}$.

Finally, we can explore the effects of the brane cosmological expansion on the dynamics of the massive mode. Assuming that $\tilde{\beta}=\tilde{\beta}(t)$ in $m^{2}=0$, the Eq.\eqref{longitudinalmodeonbranespace} leads to
\begin{equation} \label{atrator}
\ddot{\tilde{\beta}}+3H\dot{\tilde{\beta}} + 4\lambda \tilde{b}^{2}\tilde{\beta} \approx 0,
\end{equation}
where the dot is the derivative with respect to time and $H = \frac{\dot{a}(t)}{a(t)} $ is the Hubble factor. We can see again that the terms of the kinetic part of $\beta$ that depend on $z$ cancel out with the mass terms due to the constraint (\ref{law}). Note also that the cosmological expansion produces a dissipative term proportional to $3H$. For an accelerated de Sitter phase, i.e., $a(t) \propto e^{H_{0}t}$,  the solution of equation (\ref{atrator}) is given by
\begin{equation}
\tilde{\beta} = \beta_{0}e^{-\frac{1}{2}\bigg(  3H_{0} + \sqrt{9H_{0}^{2}-16\lambda\tilde{b}^{2}} \bigg)t}
\end{equation}
where $\beta_{0}$ and $H_{0}$ are constants. Assuming $H_{0}\approx 10^{16} GeV$ (inflation era), the longitudinal mode decays in a damping time $\Delta t \approx 10^{-16}(GeV)^{-1}$, corresponding to a cosmic time $~10^{-38}$ seconds.  For $m^{2}_{\beta}=\lambda\tilde{b}^2 \sim H_{0}^{2}$, the longitudinal mode has the same order as the GUT scale and it decays exponentially in time. For $m_{\beta}^{2} << H_{0}^{2}$, $\beta$ decays exponentially.
On the other hand, for
$m^{2}_{\beta}> \frac{9}{16}H_{0}^{2}$ the massive mode exhibits a damped oscillation with frequency $\omega_\beta = \sqrt{16m_{\beta}^{2}-9H_{0}^{2}}$.


\subsection{Timelike vev}

Now let us consider a timelike VEV on the 3-brane, i.e.,
\begin{equation} \label{vevbrane}
b_{M} = (\overline{b}(l/z),\vec{0},0),
\end{equation}
where $\overline{b}$ is a constant. This VEV configuration has a vanishing VEV field stregth, $ b_{MN} \neq 0 $. The transverse mode satisfies $ A^{0} \approx 0$. 

The linearized Lagrangian for this timelike VEV is given by 
\begin{eqnarray} \label{actime} \nonumber
e\mathcal{\tilde{L}}_{KS} & \approx   - \frac{1}{4} e F^{MN} F _{MN} - \frac{1}{4} e b^{MN} b_{MN}  - \frac{1}{2}e (\partial_{i}\beta)(\partial^{i}\beta)  - \frac{1}{2}e (\partial_{4}\beta)(\partial^{4}\beta) \\ 
&+e(\partial^{4}b_{0})(\partial^{0}A_{4}) - e(\partial^{4}b_{0})(\partial_{4}(\beta\hat{b}^{0})) + e(\partial_{0}A_{N})(\partial^{N}\beta) \hat{b}^{0} - 2e\lambda\overline{b}^{2}\beta^{2}.
\end{eqnarray}
The term $- \frac{1}{4} e b^{MN} b_{MN}$ will act as a source for the transverse mode, as we will see later.

We adopt a KK decomposition for the modes in order to decouple them. Assuming that  $\beta = \tilde{\beta}(x^1,x^2,x^3) \varpi(t) \Upsilon(z)$, the Eq. (\ref{motionmode}) leads to 
\begin{align} \nonumber 
  &- \bigg[ \frac{1}{(l/z)^{2}a^{3}}\partial_{0}(a^{3}\dot{\varpi}) - \bigg(\frac{4}{l^{2}} - \frac{3\ddot{a}}{a(l/z)^{2}} + \hat{b}^0\Box \hat{b}_{0}\bigg) \varpi \bigg] \frac{(l/z)^{2}}{\varpi} +  \frac{1}{\tilde{\beta}}\partial^{i}\partial_{i}\tilde{\beta}  \\ \label{sou}
  &+ \frac{1}{(l/z)^{3}\Upsilon}\partial_{4}[(l/z)^{3}\Upsilon '] + 4 \lambda \overline{b}^{2}(l/z)^{2} = \frac{(l/z)^{2}}{\tilde{\beta}\varpi\Upsilon}\bigg( \hat{b}^{0}D_{0}D_{N}A^{N} + (D_{N}b^{0N})\hat{b}_{0}\bigg) .
\end{align}
If we consider the conservation law Eq.(\ref{law}) for the timelike case, we found that
\begin{equation}
    \frac{\overline{b}\tilde{\beta} \Upsilon}{(l/z)a^{3}}\partial_{0}\bigg( a^{3}\varpi \bigg) = 0.
\end{equation} Thus, we have that $ \varpi(t) = \varpi_{0}/a^{3}$, where the $\varpi_{0} $ is a constant. Substituting this solution in Eq.(\ref{sou}), we can see that the first term vanishes. Even so, the modes are still tightly coupled. One possible setting for decoupling modes is assuming that the right side of Eq.(\ref{sou}) is zero, i.e.,  $D^{0}D_{N}A^{N}=\frac{3\overline{b}}{l^{2}(l/z)}$. Substituting this relation in Eq.(\ref{massless}), we find that the EoM for the transverse mode is given by
\begin{equation} \label{massless1}
D_{M}F^{MN}= j^{N},
\end{equation}
where 
\begin{equation}
    j^{N} = \bigg( \frac{3\overline{b}}{l^{2}(l/z)}, \vec{0}, \frac{3\overline{b}H(t)}{l(l/z)^{2}}     \bigg).
\end{equation} Thus, a source for transverse arises, since the VEV field strength $b_{MN}$ is not vanish. Assuming that $A_{\mu}(x,z)=\tilde{A}_{\mu}(x)\Gamma(z)$ and $A_{4}=0$ in Eq(\ref{massless1}), we find that
\begin{equation}\label{kk11}
    \frac{1}{\sqrt{-g_{4}}}\partial _{\mu}(\sqrt{-g_{4}}g^{\mu\alpha}_{4}g^{ij}_{4}\tilde{F}_{\alpha j}) = m^{2}\tilde{A}^{\nu}
\end{equation}
 and $\Gamma(z)$ is given by Eq.(\ref{kk2}).
Again, we notice that the location of the $A^{M}$ field in the brane with $m^{2}=0$ occurs with the help of the dilaton field.
Furthermore, from Eq. (\ref{sou}) the EoM for the longitudinal mode are 
\begin{equation} \label{ll}
    \partial^{i}\partial_{i}\tilde{\beta} = - \tilde{m}^{2}\tilde{\beta}
\end{equation}
and
\begin{equation} \label{l}
    \Upsilon '' - \frac{3}{z} \Upsilon ' + \bigg( 4\lambda \overline{b}^{2} (l/z)^{2} -\tilde{m}^{2}\bigg) \Upsilon =0.
\end{equation} For the massless longitudinal mode, i.e., $\tilde{m}^{2}=0$,  the solutions of the two equations above are, respectively, $\Upsilon(z)=\Upsilon_{0}z^{2(1\pm\sqrt{1-\lambda\overline{b}^{2}l^{2}})}$, where $\Upsilon_{0}$ is a constant, and $\tilde{\beta}$ is solution of a Laplace's equation. Let's assume that $ \lambda\overline{b}^{2}l^2 << 1$, i.e., the LV mass is small, so that $\sqrt{1 - \lambda\overline{b}^{2}l^2} \approx 1 - \frac{\lambda\overline{b}^{2}l^2}{2}$ and that $\Upsilon$ vanishes at infinity, thus it is possible to find that
\begin{equation}
    \Upsilon(z) = \Upsilon_{0}z^{-\lambda\overline{b}^{2}l^2}.
\end{equation}

Once we find solutions for massless  mode of longitudinal field through KK decomposition, we need to analyze the location of the fields in the brane again. For the longitudinal field $\beta$, we have that $S_{\parallel}= -\frac{l^4\Upsilon_{0}^{2}}{2}\int e^{-\frac{\zeta \pi (z)}{2}}z^{\pm 2\sqrt{1 - \lambda\overline{b}^{2}l^2}}$ 
$\bigg( \int \partial^{i}\tilde{\beta}\partial_{i}\tilde{\beta}\sqrt{-g_{4}}d^{4} x - \frac{4(1\pm\sqrt{1 - \lambda\overline{b}^{2}l^2})^{2}}{z^{2}}\int \tilde{\beta}^{2}\sqrt{-g_{4}}d^{4}x \bigg)dz$. 

\section{Final remarks and perspectives} \label{sec4}
We investigated how the curvature of spacetime modifies the fluctuations of a self-interacting vector field that undergoes a spontaneous Lorentz symmetry breaking. 
By considering a spacelike extra dimension and a warped geometry with a bulk cosmological constant, the bumblebee self-interaction constant $\lambda$ varies along the extra dimension.

Assuming a two parallel brane embedded in a $AdS_5$ bulk (RS-I model), the curved spacetime leads to an exponential suppression of the $\lambda$ between the branes. In the conformal coordinate (Poincar\'{e} patch), the bumblebee VEV $b_M$ also decays with the extra dimension. Therefore, the $AdS_5$ curvature of RS-I model might explain the yet unobserved massive longitudinal mode. The parameter $\alpha$ plays the role of a specific dilaton configuration. A detailed analysis of a dilaton-bumblebee action and their respective Kaluza-Klein (KK) states seems promising.

The curvature and the varying VEV turn  
the transverse NG $A_M$ and longitudinal $\beta$ modes highly coupled. Assuming the Kaluza-Klein decomposition for the modes, we find a KK mass tower for the transverse mode. The longitudinal mode only acquired a Lorentz violating mass, $m^{2}_{\beta}=\lambda b^{2}$, for the current conservation law prevents $\beta$ to acquire KK masses.
The brane curvature due to the cosmological expansion leads to a dissipative term proportional to the Hubble constant. For a De Sitter accelerated expansion, the time decay is proportional to $1/H_{0}$. Thus, the cosmic expansion dilutes the longitudinal mode leaving only the NG modes in late times.

For a timelike VEV, the longitudinal mode decouples from the NG modes and it is not propagating, as in the Minkowski \cite{ks,ks2}. In addition, assuming a time-dependent amplitude, the massive mode decays with $a^{-3}$. Therefore, if the spontaneous violation of the Lorentz symmetry occurred in the early universe, the inflationary period may have strongly suppressed the effects of the longitudinal mode. This result suggests further analysis on the effects of combined bumblebee, gravity and matter fluctuation effects in the early universe. 

\section*{Acknowledgments}
C.A.S. Almeida thanks the Conselho Nacional de Desenvolvimento Cient\'{\i}fico e Tecnol\'{o}gico (CNPq), grant 
$n\textsuperscript{\underline{\scriptsize o}}$ $308638/2015-8$ for financial support.


\end{document}